\documentstyle[11pt,psfig]{article}
\newcommand{\sspc}{\hspace{0.1in}}
\newcommand{\spc}{\hspace{0.22in}}

\newcommand{\spn}{\hspace*{0.25in}}

\def \ds {\displaystyle}

\def \sp {\scriptsize}
\def \ts {\thinspace}

\begin{document}
\begin{large}
%
%
\begin{center}
\hspace*{7.5cm} {\bf KEK-CP-040} \\
\hspace*{6.7cm} {\bf UTPP-45} \\
\vspace{1.0cm}
{\LARGE {\bf Phases and fractal structures}} \\
\vspace{0.3cm}
{\LARGE {\bf of three-dimensional simplicial gravity}} \\
\vspace{1.5cm}
{\large {\sc Hiroyuki Hagura}}\footnote{1 e-mail: hagura@theory.kek.jp}  \\
\vspace{0.3cm}
{\em Institute of Physics, University of Tsukuba, \\
                        Tsukuba-shi Ibaraki 305, Japan} \\
\vspace{0.6cm}
{\large {\sc Noritsugu Tsuda}}\footnote{2 e-mail: ntsuda@theory.kek.jp} \\
\vspace{0.3cm}
{\em National Laboratory for High Energy Physics} (KEK), \\
                        {\em Tsukuba-shi Ibaraki 305, Japan} \\
\vspace{0.6cm}
{\large {\sc Tetsuyuki Yukawa}}\footnote{3 e-mail: yukawa@theory.kek.jp} \\
\vspace{0.3cm}
{\em National Laboratory for High Energy Physics} (KEK), \\ and \\
{\em Coordination Center for Research and Education, \\
     The Graduate University for Advanced Studies,} \\
     {\em Kamiyamaguchi Hayama-chyo Miura-gun Kanagawa 240-01, Japan}
\end{center}
\vspace{0.3cm}
%
%
\begin{abstract}
\sspc We study phases and fractal structures of three-dimensional
simplicial quantum gravity by the Monte-Carlo method. After measuring the
surface area distribution (SAD) which is the three-dimensional analog of
the loop length distribution (LLD) in two-dimensional quantum gravity, we
classify the fractal structures into three types:
(i) in the hot (strong coupling) phase, strong gravity makes the space-time one
crumpled mother universe with small fluctuating branches around it. This is a
crumpled phase with a large Hausdorff dimension $d_{\mbox{\tiny H}} \simeq 5$.
The topologies of cross-sections are extremely complicated.
(ii) at the critical point, we observe that the space-time is a fractal-like
manifold which has one mother universe with small and middle size branches
around it. The Hausdorff dimension is $d_{\mbox{\tiny H}} \simeq 4$.
We observe some scaling behaviors for the cross-sections of the
manifold. This manifold resembles the fractal surface observed in
two-dimensional quantum gravity.
(iii) in the cold (weak coupling) phase, the mother universe disappears
completely and the space-time seems to be the branched-polymer with a small
Hausdorff dimension $d_{\mbox{\tiny H}} \simeq 2$. Almost all of the
cross-sections have the spherical topology $S^2$ in the cold phase.
\end{abstract}
\vspace{1cm}
\newpage
%
%
\section{Introduction}

\spc In recent years remarkable progress has been made to quantize the field
theories with general covariance . Especially, in two-dimensional quantum
gravity it is very encouraging
that some analytical formulations (Liouville field theory\cite{DK}, matrix
models\cite{GM}, topological field theories\cite{DVV}\cite{BBRT})
and numerical calculations by the dynamical triangulations
(DT)\cite{ADF}\cite{BKKM}\cite{JKPS} have shown the good agreement. It is found
that the universal structure which leads to the continuum limit seems to be
its fractal structure\cite{KKMW}. It seems very important to investigate
the fractal behavior of the space-time manifold in any dimension. Below,
we will briefly review this fractal structure in two dimension as the
starting point.

  On the other hand, in the three-dimensional quantum gravity which is our
primary concern of this paper, as well as in the
four-dimensional case, we have no analytical models which predict the scaling
or fractal properties near the critical point. By numerical simulations, a few
groups have studied the three-dimensional Euclidean quantum gravity
based on dynamical triangulations and reported the existence of the first-order
phase transition\cite{ABKV}\cite{AVT}. Numerical calculations show that the
entropy is exponentially bounded and the thermodynamic limit exists in
three-dimensional simplicial gravity\cite{AVE}\cite{CKR}. The starting point
of this model is based on the ansatz that the partition function
describing the fluctuations of a continuum geometry can be approximated by
a weighted sum over all simplicial manifolds (triangulations) $T$ which is
composed of equilateral tetrahedra:
\begin{eqnarray}
Z = \sum_{T} e^{-S_0(T)} \ts.
\end{eqnarray}
In this work the topology of the simplicial lattice is restricted to the
three-sphere $S^3$. The action is taken to be of the form
\begin{eqnarray}
S_0(T) = - \kappa_0 N_0 + \kappa_3 N_3 \ts,
\label{eq:S_0}
\end{eqnarray}
where $N_i$ is the number of $i$-simplices in a simplicial lattice $T$, and
$\kappa_0$, $\kappa_3$ are dimensionless gravitational and cosmological
constants on the lattice.  This action (\ref{eq:S_0}) corresponds to a
discretized version of the continuum action of three-dimensional
Einstein gravity:
\begin{eqnarray}
S_{\mbox{\sp EH}}[g] = - \frac{\ds 1}{\ds G} \int d^3 x \sqrt{g}R
                       + \lambda\int d^3 x \sqrt{g} \enspace,
\label{eq:S_C}
\end{eqnarray}
where $G$ is the Newton constant and $\lambda$ is the cosmological constant.

   To control the volume fluctuations we add an additional term to the action
of the form:
\begin{eqnarray}
\delta S = \gamma \left(N_3 - V\right)^2 \enspace ,
\label{eq:delta_S}
\end{eqnarray}
where $V$ is the target size of the simplicial lattices and $\gamma$ is an
appropriate parameter. In ref.\cite{CKR} it is argued that lattices
with $N_3 \sim V$ are
distributed according to the correct Boltzmann weight up to correction terms of
order $O(\frac{1}{\sqrt{\gamma}V})$.

   In summary, our starting point is the following partition function:
\begin{eqnarray}
Z = \sum_{T} e^{-S(T)} \enspace,
\label{eq:full_PF}
\end{eqnarray}
where the total action $S(T)$ on the three-dimensional simplicial lattice is
of the form
\begin{eqnarray}
S(T) = - \kappa_0 N_0 + \kappa_3 N_3 + \gamma \left(N_3 - V\right)^2 \enspace.
\label{eq:full_S}
\end{eqnarray}

   Now, we would like to explain the method to detect the fractal structure,
taking two-dimensional quantum gravity as an example in some detail.
Let us consider a triangulated surface $\mit\Sigma$ which is composed of
equilateral triangles.
One of the points to understand the continuum limit of the DT in
two dimension is its fractal structure\cite{KKMW}, characterized by the
loop length distribution (LLD) defined as follows:
\begin{enumerate}
\item Let us consider a $\phi^3$-graph $\tilde{\mit\Sigma}$ dual to the
triangulated surface $\mit\Sigma$.
\item Pick up a single point $P$ in $\tilde{\mit\Sigma}$
(a triangle in $\mit\Sigma$).
\item Find all the points which has a geodesic distance (= minimum number of
steps of two points in the dual graph) $D$ from $P$.
\item The set of links (also links in $\mit\Sigma$) connecting two points of
distance $D-1$ and $D$ consists of some closed loops in $\mit\Sigma$.
\item We define the {\em loop length distribution} (LLD)  function by counting
the number of these closed loops which have a geodesic distance $D$ and a loop
length $L$, and write $g(L, D)$.
\end{enumerate}
As shown in ref.\cite{KKMW}, the function $g(L, D)$ is universal and
explicitly calculated as
\begin{eqnarray}
g(L, D) \times D^2 = \frac{\ds 3}{\ds 7\sqrt{\pi}} ( x^{-5/2} + \frac{\ds 1}
{\ds 2}x^{-3/2} + \frac{\ds 14}{\ds 3}x^{1/2} ) e^{-x}
\end{eqnarray}
with $x = L/D^2$ as a scaling parameter. Numerically, this fractal behavior
was established by N.Tsuda and T.Yukawa\cite{TY}\cite{TYRS}.

   In the next section, we will explain our data and the interpretations of
the measurements will be described, mainly on the three-dimensional analog
of the fractal structure described above.
\vspace{0.5cm}
%
%
\section{Numerical simulations}
\spn To discuss the phase structure of simplicial gravity, the integrated
scalar curvature per volume has conventionally been used.
The lattice analog is:
\begin{eqnarray}
\frac{\ds \int d^3 x \sqrt{g}R}{\ds \int d^3 x \sqrt{g}}
\quad \longrightarrow \quad
\sum_l \frac{\ds \left( c - n(l) \right)}{\ds N_3}
= c\frac{\ds N_0}{\ds N_3} - (6-c)\enspace ,
\label{eq:ord_para}
\end{eqnarray}
where $c = 2\pi / \arccos(1/3) = 5.104\dots$ is the number of tetrahedra
around a link which gives the zero scalar curvature for flat space.
Here $n(l)$ denotes the number of tetrahedra sharing the link $l$, called the
{\em coordination number}.
A conventional indicator of phase transitions is the specific heat, which in
this case corresponds to the fluctuation in the number of vertices $N_0$:
\begin{eqnarray}
\chi_{N_0} = \frac{\ds 1}{\ds V} (\langle N_0^2\rangle - \langle N_0\rangle^2 )
\enspace ,
\label{eq:chi_N0}
\end{eqnarray}
where averages are taken over ensembles of configurations with a fixed volume
$N_3 = V$.
In fig.\ref{fig:hyst}, we show the expectation value $<N_0/N_3>$ and
$\chi_{N_0}$ as a function of the gravitational coupling constant $\kappa_0$.
We cannot observe the hysteresis curve, in contrast to the calculation in
ref.\cite{ABKV}. We observe a phase transition near $\kappa_0 = 4.0$,
but with this data above, we cannot determine the order of the transition
unless performing the finite-size scaling analysis.

   Here we should make a remark about the thermalization in our study which
will explain the difference from the paper\cite{ABKV}.
For the error estimate we used the jackknife method\cite{JK}. We examined the
magnitude of error as a function of the bin size and adopted the bin size where
the jackknife error had leveled off. This bin size corresponds to about 500
sweeps where one sweep is defined as the total number of accepted moves divided
by the number of tetrahedra $N_3$. Comparing to the typical interval of about
100 sweeps in two dimension, the relaxation time of the Markov chain in three
dimension is much longer. We need to perform computer runs with the long
interval larger than 500 sweeps to accumulate the statistically uncorrelated
data. We have taken the interval 600 sweeps to calculate our data.
The reason why the hysteresis has disappeared in fig.\ref{fig:hyst} is that we
have thermalized our data enough to avoid the autocorrelation. In
ref.\cite{ABKV}, $10^{3}$ sweeps were taken to get each point in the
hysteresis, but we have taken $600 \times 120 = 7.2 \times 10^{4}$ sweeps to
get each point in the fig.\ref{fig:hyst}. No hysteresis is often observed
in a finite system,
though there exists a first-order transition in the system\cite{FMOU}. We
cannot determine the order of the transition without performing the
finite-size scaling analysis\cite{CLB}.

   In general, if a statistical system has a
second-order phase transition, some scaling invariant properties, or, fractal
structures are expected to appear at the critical point. In the case of
two-dimensional gravity, the LLD characterizes the fractal property of the
surface\cite{KKMW} as described in the last section.
By the analogy to two-dimensional case, we introduce the surface area
distribution (SAD) in three-dimensional case:
\begin{enumerate}
\item Let us consider a connected $\phi^4$-graph $\tilde{T}$ dual to a
three-dimensional simplicial lattice $T$ with a spherical topology $S^3$.
\item Pick up a single point $P$ in $\tilde{T}$ (a tetrahedron in $T$).
\item Find all the points which has a geodesic distance (= minimum number of
steps of two points in the dual graph) $D$ from $P$.
\item The set of links (triangles in $T$) connecting two points of
distance $D-1$ and $D$ consist of some closed surfaces in $T$.
\item We define the distribution function of these closed surfaces which have
a distance $D$ and area $A$ as {\em surface area distribution} (SAD), and
denote by $f(A, D)$.
\end{enumerate}
Theoretically, there exist no analytical models which predict the scaling
behavior of this SAD function $f(A, D)$. We have measured the SAD functions in
the case of the lattice size $V = 10000$. At each distance $D$, we compute the
number of branches (baby universes) $N_{\mbox{\sp b}}(D)$, the volume within
the distance $V(D)$ , and moreover the large area ($\equiv$ the
cross-section of the mother universe) $A_{\mbox{\sp large}}(D)$.
In general, we expect these observables to scale as,
\begin{eqnarray}
N_{\mbox{\sp b}}(D)     &\sim& D^{d_{\mbox{\tiny b}}} \ts, \nonumber \\
V(D) \hspace*{0.01cm}   &\sim& D^{d_{\mbox{\tiny H}}} \ts, \\
A_{\mbox{\sp large}}(D) &\sim& D^{d_{\mbox{\tiny M}}} \ts. \nonumber
\label{eq:HM_scale}
\end{eqnarray}
If the manifold (space-time) has a mother body (huge universe) which
dominates the fractal structure, we expect
$d_{\mbox{\sp H}} = d_{\mbox{\sp M}} + 1$.
\vspace{0.5cm}

   In fig.\ref{fig:SAD_hot}, we show the data in the strong coupling limit
$\kappa_0 = 0$. There exists a huge mother universe.
The behavior of $N_{\mbox{\sp b}}(D)$ in fig.\ref{fig:SAD_hot}(a) is similar
to that of the crumpled surface in two-dimensional quantum gravity\cite{TYRS}.
{}From the linear region $D \simeq 8$
in fig.\ref{fig:SAD_hot}(b),
we can read $V(D) \sim D^5$, or the Hausdorff dimension $d_{\mbox{\sp H}}
\simeq 5$.
In fig.\ref{fig:SAD_hot}(c), we show the SAD with $x = A/D^4$ as a scaling
variable, because this variable gives the most proper fitting for the
distribution of the large area ($\equiv$ cross-section of the mother universe).
Fig.\ref{fig:SAD_hot}(d) also shows that the large area should scale as
$A_{\mbox{\sp large}}(D) \sim D^4$, consistent with the naive expectation
$d_{\mbox{\sp H}} = d_{\mbox{\sp M}} + 1$.
In fig.\ref{fig:SAD_hot}(c), we can observe no contributions from the universes
of middle size. These facts give us the picture that the space-time manifold
has one extremely crumpled huge mother universe which is not expanding widely
and many small baby universes of low density are fluctuating around it.
The SAD function for the mother universe which dominates the fractal
structure of the manifold seems to indicate a scaling behavior with a scaling
variable $x = A/D^4$.
\vspace{0.5cm}

   Next, the data near the critical point $\kappa_0 = 4.0$ are shown in
fig.\ref{fig:SAD_crit}. The behavior of $N_{\mbox{\sp b}}(D)$ in
fig.\ref{fig:SAD_crit}(a) is very similar to that of the fractal surface in
two-dimensional quantum gravity\cite{TYRS}. This is one of the reasons why
we call the manifold at the critical point {\em fractal-like}. We can read
$d_{\mbox{\sp b}} \simeq 3$ from fig.\ref{fig:SAD_crit}(a).
{}From the linear region in fig.\ref{fig:SAD_crit}(b), we can read the
Hausdorff
dimension $d_{\mbox{\sp H}} \simeq 4$.
Our data show the existence of the mother universe, which extends more widely
than that of the hot phase. Fig.\ref{fig:SAD_crit}(d) shows that the large
area ($\equiv$ cross-section of the mother universe) scales as
$A_{\mbox{\sp large}} \sim D^3$, giving the same Hausdorff dimension
$d_{\mbox{\sp H}} \simeq 4$. So we choose $x = A/D^3$ as the scaling variable,
but the small and middle areas ($\equiv$ rest parts after subtracting the
contribution of the mother) do not scale as well by this variable in
fig.\ref{fig:SAD_crit}(c).
Then, we divide the SAD function into two parts, the contribution of the large
area $f_{\mbox{\scriptsize large}}(A,\ts D)$ and rest of areas
(hereafter we abbreviate small and middle areas to small areas)
$f_{\mbox{\scriptsize small}}(A,\ts D) \ts$:
\begin{eqnarray}
f(A,\ts D) = f_{\mbox{\sp large}}(A,\ts D) + f_{\mbox{\sp small}}(A,\ts D) \ts.
\label{eq:f_divide}
\end{eqnarray}
The divided data in fig.\ref{fig:SAD_crit_msd}(b) tells us that
$x^{\prime} = A/D^2$ is a good scaling variable for the dominant part of
$f_{\mbox{\sp small}}(A,\ts D)$.

   Let us discuss the fractal(-like) structure at the critical point in some
detail. In fig.\ref{fig:SAD_crit_msd}(a), we can assume the large area
distribution function as,
\begin{eqnarray}
f_{\mbox{\scriptsize large}}(A,\ts D) = const. \times \frac{\ds 1}{\ds  D^3}
\ts x^{a} e^{-bx} \ts .
\label{eq:f_large}
\end{eqnarray}
Here $a$ and $b$ are some constants. The fittings in the
fig.\ref{fig:SAD_crit_msd}(a) show that we can safely assume $a > -1$, so the
integration of $f_{\mbox{\scriptsize large}}(A,\ts D)$ given in
eq.(\ref{eq:f_large}) is convergent. In other words, the following quantities
are finite:
\begin{eqnarray}
\langle A^{n} \rangle_{\mbox{\sp large}} = \ts \lim_{\epsilon \to 0}
\int^{\infty}_{\epsilon} dA \ts
 A^n f_{\mbox{\scriptsize large}}(A,\ts D) \hspace*{1cm} (n \geq 0) \ts .
\label{eq:A_large^n}
\end{eqnarray}
Here $\epsilon$ is the cutoff for the area on the simplicial lattice.
It is easy to show that
\begin{eqnarray}
\langle A^0 \rangle_{\mbox{\sp large}}
&\propto& \mit\Gamma(a + \mbox{1}) \times D^{\mbox{\sp 0}} \ts , \nonumber\\
\langle A^1 \rangle_{\mbox{\sp large}}
&\propto& \mit\Gamma(a + \mbox{2}) \times D^{\mbox{\sp 3}} \ts .
\label{eq:A_large}
\end{eqnarray}
{}From eq.(\ref{eq:A_large}), the large area part (mother universe) is
independent of the cutoff $\epsilon$, so this part seems to be the universal
quantity.

   On the other hand, from fig.\ref{fig:SAD_crit_msd}(b) the small area
distribution function is parametrized as,
\begin{eqnarray}
f_{\mbox{\sp small}}(A,\ts D) = const. \times \frac{\ds 1}{\ds  D^2} \ts
x^{\prime -a^{\prime}} e^{-b^{\prime} x^{\prime}} \ts .
\label{eq:f_small}
\end{eqnarray}
Here $a^{\prime}$ and $b^{\prime}$ are positive constants. From the gradient of
the dominant line in the fig.\ref{fig:SAD_crit_msd}(b), we can read the value
of the exponent $a^{\prime} = 2/5$. The good fitting function (after rescaling
$b^{\prime} x^{\prime} \to x^{\prime}$) is the following form:
\begin{eqnarray}
f_{\mbox{\sp small}}(A,\ts D) = const. \times \frac{\ds 1}{\ds  D^2} \ts
x^{\prime -5/2} e^{-x^{\prime}} \ts .
\label{eq:f_small_fit}
\end{eqnarray}
Then, we can calculate the quantity:
\begin{eqnarray}
\langle A^0 \rangle_{\mbox{\sp small}}
&   =   &\int^{\infty}_{\epsilon}
dA \ts f_{\mbox{\sp small}}(A,\ts D) \nonumber \\
&\propto&  const.\times D^3 \epsilon^{-3/2} + const.\times D \epsilon^{-1/2}
+ const. \times D^0 \ts ,
\label{eq:A_small^0} \\
\langle A^1 \rangle_{\mbox{\sp small}}
&   =   &\int^{\infty}_{\epsilon}
dA \ts A \ts f_{\mbox{\sp small}}(A,\ts D) \nonumber \\
&\propto&  const. \times D^3 \epsilon^{-1/2} + const. \times D^2 \ts .
\label{eq:A_small}
\end{eqnarray}
Moreover, the average area of the small part behaves as
\begin{eqnarray}
\langle A^1 \rangle_{\mbox{\sp small}} / \langle A^0 \rangle_{\mbox{\sp small}}
\ts \sim \ts \epsilon \ts.
\label{eq:A_small_per}
\end{eqnarray}
This relation (\ref{eq:A_small_per}) is quite similar to the two-dimensional
fractal surface where the small loop has a constant length with the
size of the cutoff length\cite{KKMW}. The small baby universe is not a
universal quantity but also a local quantum fluctuation.
If we identify
\begin{eqnarray}
\langle A^0 \rangle_{\mbox{\sp small}} = N_{\mbox{\sp b}}(D) \sim
D^{d_{\mbox{\tiny b}}} \ts,
\label{eq:A_small^0&N_b}
\end{eqnarray}
then we can decide $d_{\mbox{\tiny b}} = 3$ by comparing
eq.(\ref{eq:A_small^0}) and eq.(\ref{eq:A_small^0&N_b}). This value of
$d_{\mbox{\tiny b}} = 3$ accounts very well for the behavior of
$N_{\mbox{\sp b}}(D)$ in fig.\ref{fig:SAD_crit}(a). Moreover,
eq.(\ref{eq:A_small}) and eq.(\ref{eq:A_large}) give the same Hausdorff
dimension $d_{\mbox{\sp H}} = 4$, accounting for the behavior of $V(D)$ in
fig.\ref{fig:SAD_crit}(b).

   Here we mention the observation by Agishtein and Migdal\cite{AM}.
It was found in ref.\cite{AM} that the Hausdorff dimension $d_{\mbox{\sp H}}$
grows logarithmically
\begin{eqnarray}
d_{\mbox{\sp H}} \sim c_1 + c_2 \log D  \hspace*{2cm}
(c_1, c_2 \enspace \mbox{are constants}) \ts .
\label{eq:d_H_log}
\end{eqnarray}
They argued that the logarithmic growth of Hausdorff dimension might be a
universal law of Nature. But, in our interpretation, this behavior
(\ref{eq:d_H_log}) expresses the non-universal (cutoff-dependent) property
which originates in the small part (see eq.(\ref{eq:A_small})).

  Now we can explain the notion of the {\em fractal-like manifold}. At the
critical point, the mother, middle and small universes contribute to the SAD
function as well as to the LLD function of the fractal surface in
two-dimensional quantum gravity\cite{KKMW}. But the distinct point is the
observation that we have two kinds of scaling variables, $x = A/D^3$ for
the mother and $x^{\prime} = A/D^2$ for the baby universes in
three-dimensional quantum gravity.
This is why we call such a manifold not fractal but fractal-like. It is a
remarkable property of three-dimensional fractal-like manifold that both the
large and small areas give the same Hausdorff dimension $d_{\mbox{\sp H}}
\simeq 4$, though the large area gives $d_{\mbox{\sp H}} = 3$ but the
dominant small areas give $d_{\mbox{\sp H}} = 4$ in two-dimensional fractal
manifold\cite{KKMW}.
\vspace{0.5cm}

   Finally, we show the data in the weak coupling phase (cold phase)
$\kappa_0 = 5.0$ in fig.\ref{fig:SAD_cold}. In this phase, the space-time
manifold consists of widely expanding branches and has no mother universe.
Fig.\ref{fig:SAD_cold}(a) shows $N_{\mbox{\sp b}}(D) \sim D$. The behavior of
$N_{\mbox{\sp b}}(D)$ is quite different from the other cases, but similar to
that of the branched-polymer in two-dimensional gravity\cite{TYRS}.
 From fig.\ref{fig:SAD_cold}(b), we can easily
read that $V(D) \sim D^2$, or $d_{\mbox{\sp H}} \simeq 2$.
Because the mother body does not exist in this phase,
we should choose a scaling variable $x = A/D$
for the dominant small fluctuations in fig.\ref{fig:SAD_cold}(c).
\vspace{1cm}
%
%
\section{Summary and discussion}
\spc There exist two phases, hot and cold phase, in three-dimensional
simplicial quantum gravity, as previously known\cite{BK}\cite{AVT}. They
argued the first-order transition between these two phases because of the
existence of a large hysteresis\cite{ABKV}. In our study we observed a rapid
transition near $\kappa_0 = 4.0$ but no hysteresis, so we cannot determine
the order of the transition without the finite-size scaling analysis.

   By measuring the SAD functions using the Monte-Carlo technique, we could
characterize the phases of three-dimensional simplicial gravity  into three
types:
\begin{enumerate}
\item The hot phase (strong coupling phase).\\
\hspace*{0.12in} This phase is a typical crumpled phase. Strong gravity makes
the space-time a crumpled manifold with small fluctuations around it. The fact
that the area of the cross-section of the mother universe scales as
$A_{\mbox{\sp large}}(D) \sim D^4$ determines the Hausdorff dimension to be
$d_{\mbox{\sp H}} \simeq 5$. The topology of the cross-sections for the mother
universe is highly complicated, though those of baby universes are almost
spherical topologies. Because the contribution of the baby universes to the SAD
function is quite different from that of the mother in contrast to the LLD in
two dimension, we do not know how one associates a fractal structure
with this case.
\item The critical point. \\
\hspace*{0.12in} At this point, the space-time is a fractal-like manifold.
By the term
fractal-like, we means that this manifold seems to have two different scaling
variables, $x = A/D^3$ for the mother universe, and $x^{\prime} = A/D^2$ for
the baby universes. Both the long and short distance regions give the same
Hausdorff dimension $d_{\mbox{\sp H}} \simeq 4$. The fractal behavior of the
long distance part seems to be universal or cutoff-independent.
The topologies of the cross-sections are less complicated than those in the
hot phase.
\item The cold phase (weak coupling phase). \\
\hspace*{0.12in} In this phase, the mother universe, which exists in the hot
phase and at the critical point, disappears completely. Weak gravity cannot
make any large body in the space-time. The fractal structure is determined
by the dominant small fluctuations with a scaling parameter $x = A/D$. Almost
all of the cross-sections have the spherical topology $S^2$. The space-time
is a branched-polymer with a small Hausdorff dimension $d_{\mbox{\sp H}}
\simeq 2$. We could not find any universal (cutoff-independent) property in
this phase.
\end{enumerate}

   We should discuss the properties at the critical point in more detail.
Because three-dimensional Einstein-Hilbert action (\ref{eq:S_C}) includes a
dimensionfull coupling constant $G$, this theory is perturbatively
unrenormalizable as well as four-dimensional Einstein gravity. In other words,
we cannot expect the scale invariance over all order of scales in such a
theory. But our data shows that two different types of scaling regions seem to
exist near the critical point, short distance region (small baby universe) and
long distance region (mother universe).
Short distance region has its own scaling with a scaling variable
$x^{\prime} = A/D^2$. But this region seems to be cutoff-dependent, so we
cannot expect the universal structure in this region.
On the other hand, long distance region also has its own scaling with a scaling
variable $x = A/D^3$. As the scaling of this region seems to be
cutoff-independent, this fractal-like structure may be a universal one which
leads to the continuum limit as well as the fractal structure does in
two-dimensional lattice gravity. This point is very important. To confirm our
conjecture, we added the irrelevant discretized $R^2$-term to the
Einstein-Hilbert action (\ref{eq:full_S}) on the simplicial lattice.
Though the SAD function of the small part changed its form, that of the mother
part did not change its form when the coupling to $R^2$-term is
small\cite{HYR2}. But these facts do not establish the universality of the
fractal-like SAD function, so we need further studies based on the method in
constructive field theory and interesting results are expected.
\vspace{1cm}

\noindent{{\bf Acknowledgements}} \\
\smallskip
\spc We would like to thank the members of KEK theory group for
discussions. We are especially grateful to H.Kawai and N.Ishibashi for
useful discussions and critical comments.
\vspace*{0.8cm}
%
%

%
%
\begin{figure}[h]
\vspace{1.5cm}
\centerline{\psfig{file=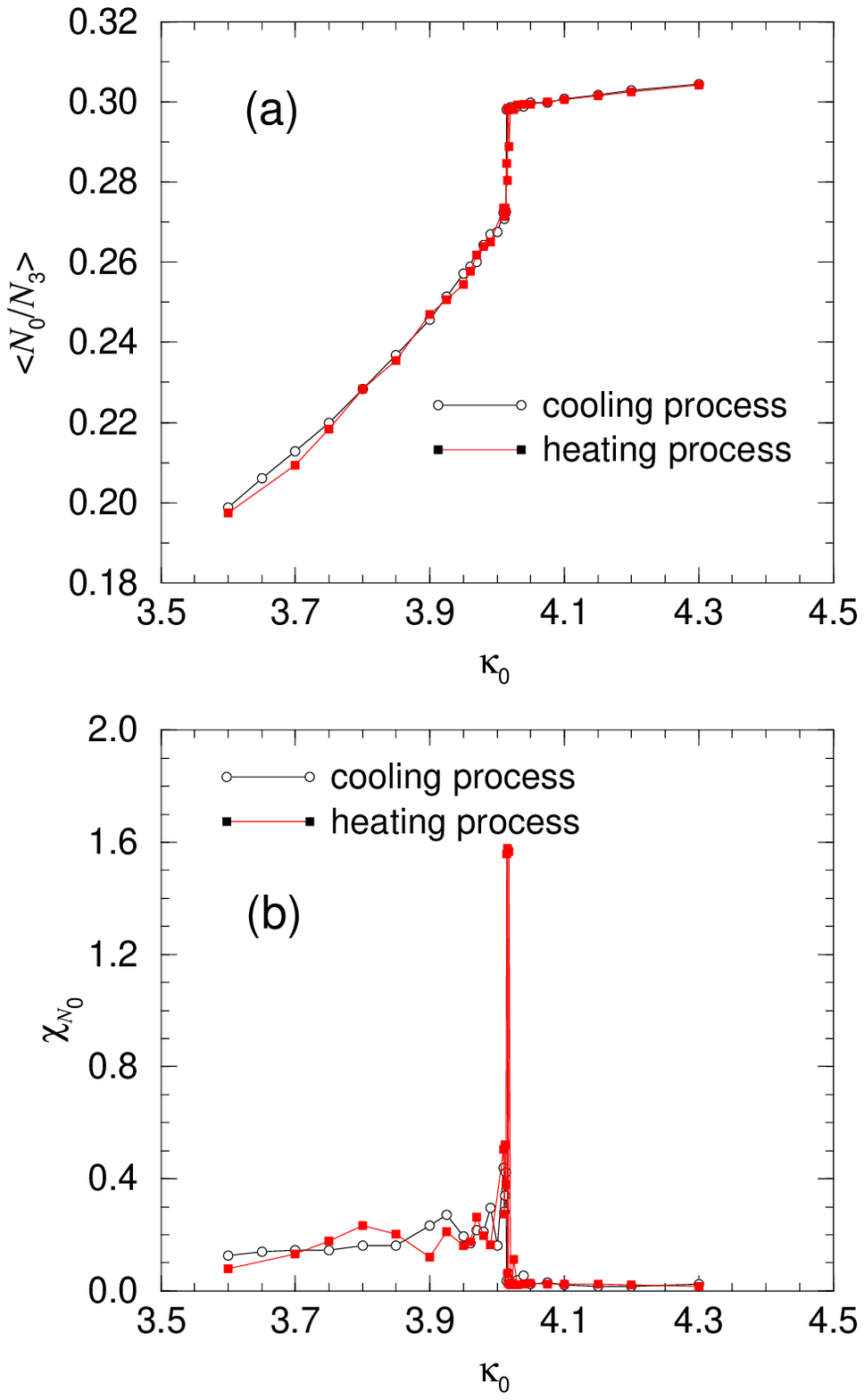,height=14cm,width=16cm}}
\caption{Cooling and heating processes with lattice size $V$ = 10000.
(a)Scalar curvature per volume .vs. $\kappa_0$ for each process. No hysteresis
is observed. (b)Node susceptibility .vs. $\kappa_0$}
\label{fig:hyst}
\end{figure}
%
%
\begin{figure}[h]
\vspace{1cm}
\centerline{\psfig{file=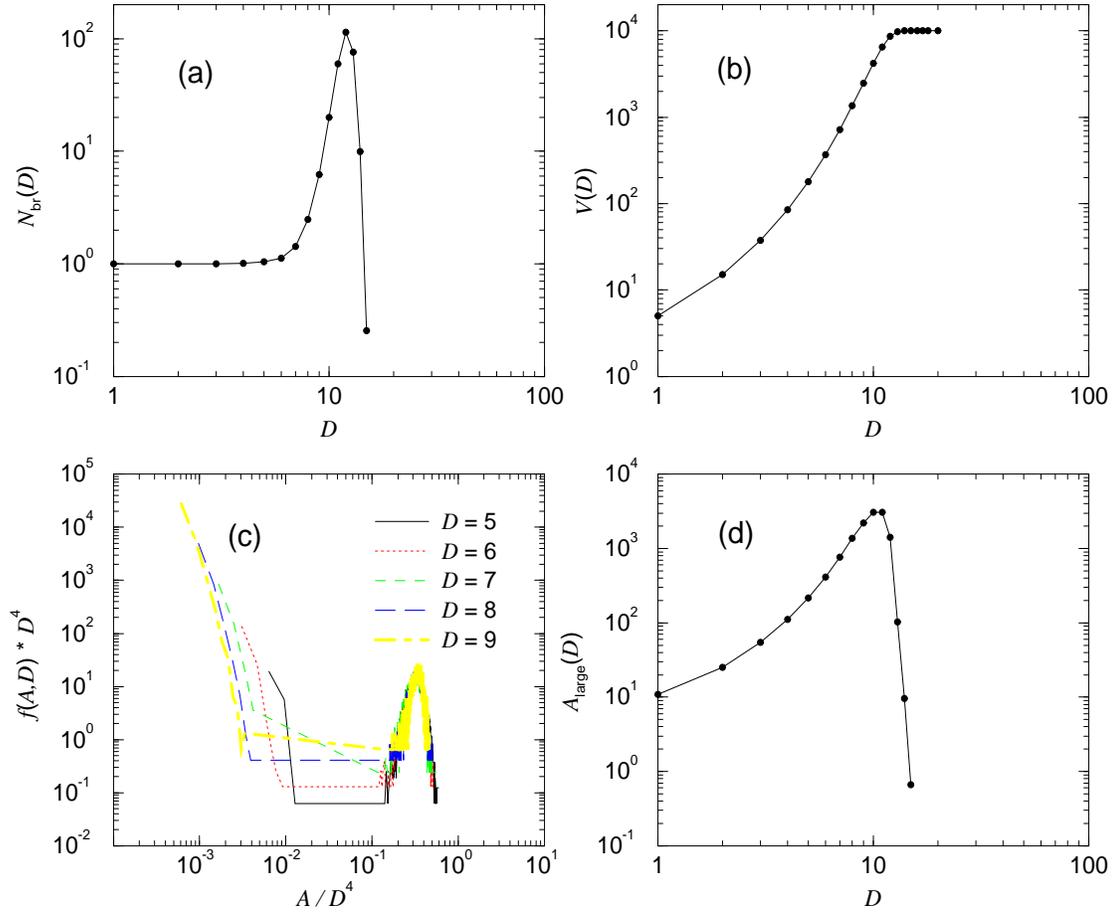,height=12cm,width=14cm}}
\caption{The data in the hot phase (strong coupling limit $\kappa_0 = 0$).
(a)Number of branches at each distance $D$. (b)Volume within the geodesic
distance $D$. (c)The SAD is shown with a scaling parameter $x = A/D^4$.
(d) The expectation value of the large area.}
\label{fig:SAD_hot}
\vspace{0.5cm}
\end{figure}
%
%
\begin{figure}[h]
\vspace{1cm}
\centerline{\psfig{file=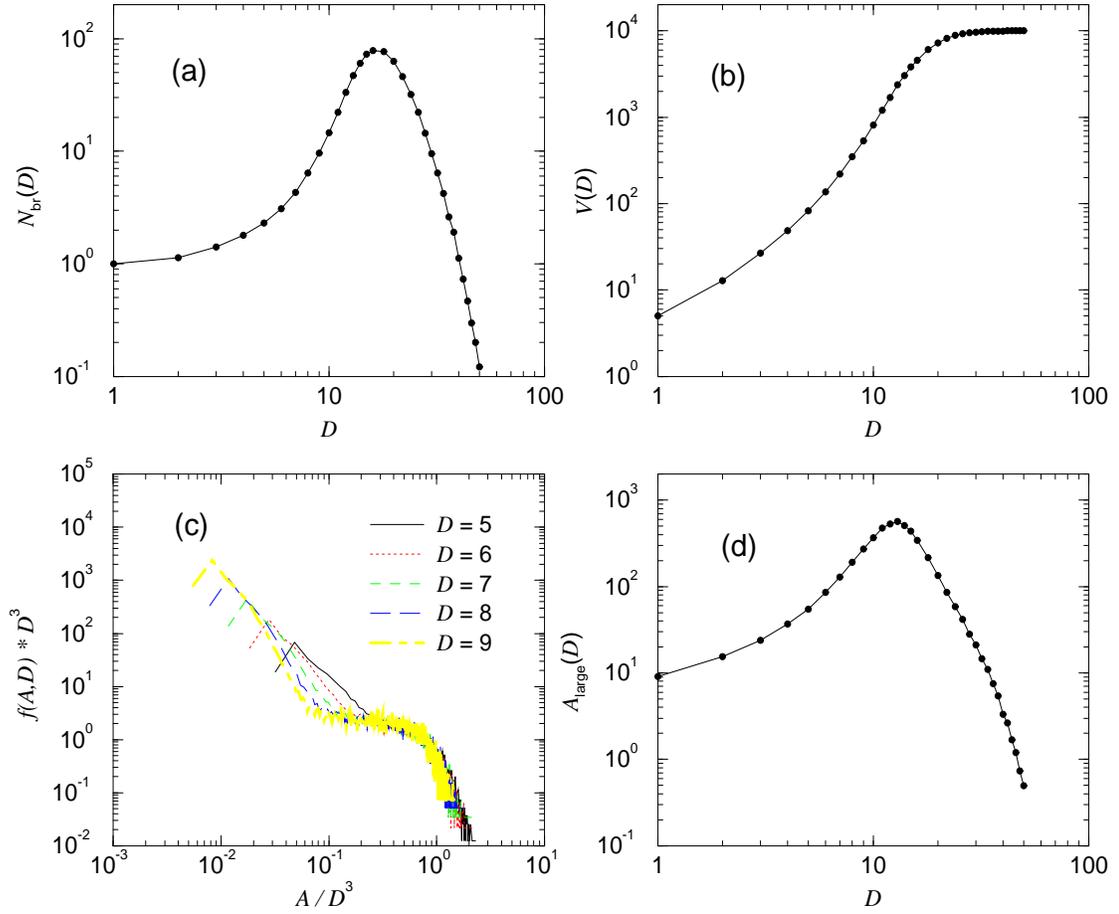,height=12cm,width=14cm}}
\caption{The data at the critical point ($\kappa_0 = 4.0$).
(a)Number of branches at each distance $D$. (b)Volume within the geodesic
distance $D$.
(c)The SAD is shown with the scaling parameter
$x = A/D^3$. (d) The expectation value of the large area.}
\label{fig:SAD_crit}
\vspace{0.5cm}
\end{figure}
\vspace{0.5cm}
%
%
\begin{figure}
\vspace{1cm}
\centerline{\psfig{file=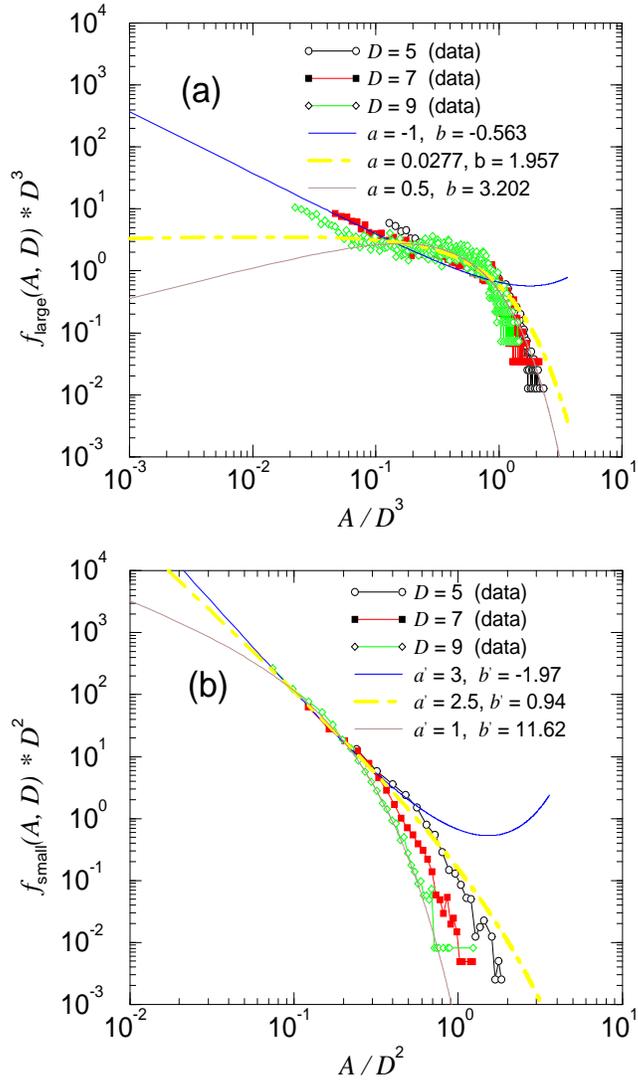,height=14cm,width=16cm}}
\caption{The divided SAD function at the critical point ($\kappa_0 = 4.0$).
(a)The SAD for the large part measured and some fittings.
(b)The SAD for the small part measured and some fittings.}
\label{fig:SAD_crit_msd}
\vspace{0.5cm}
\end{figure}
%
%
\begin{figure}[h]
\vspace{1cm}
\centerline{\psfig{file=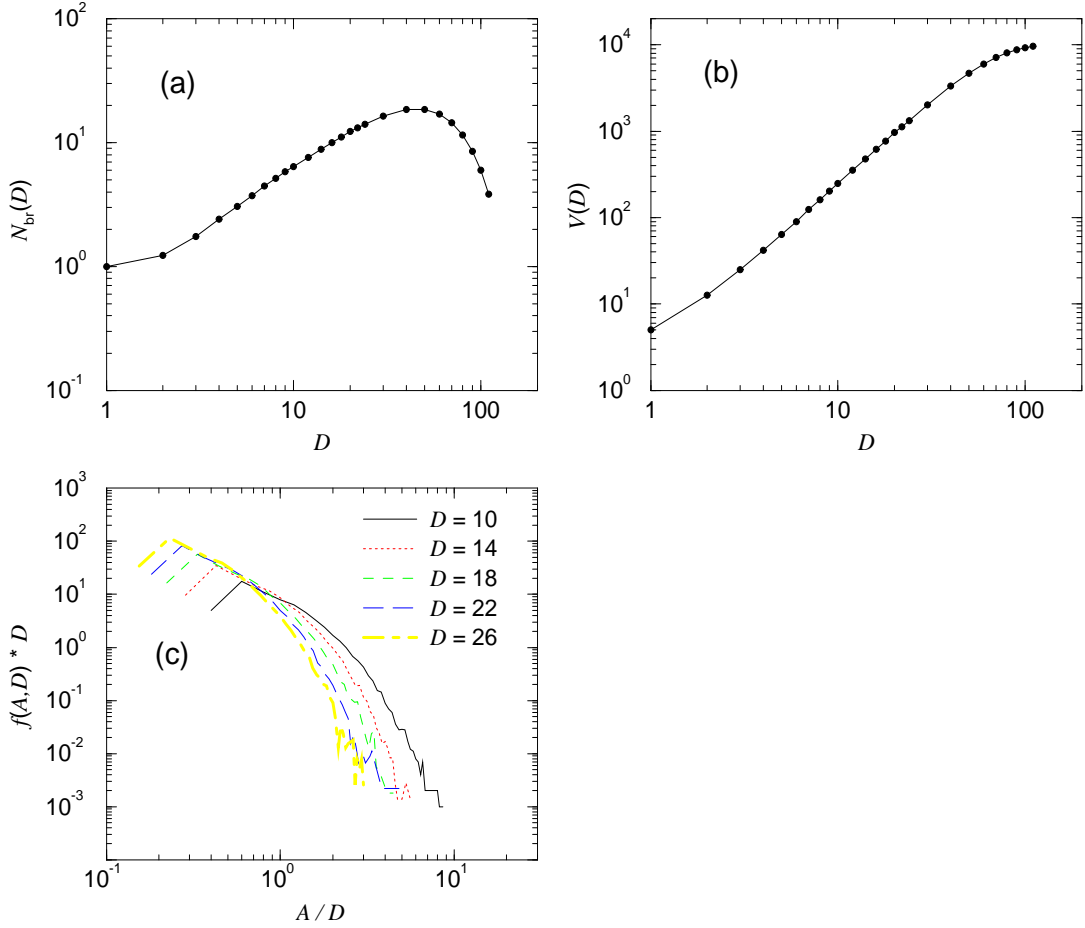,height=12cm,width=14cm}}
\caption{The data in the cold phase ($\kappa_0 = 5.0$).
(a)Number of branches at each distance $D$. (b)Volume within the geodesic
distance $D$.
(c)The SAD is shown with the scaling parameter $x = A/D$.}
\label{fig:SAD_cold}
\vspace{0.5cm}
\end{figure}
\end{large}

\begin{thebibliography}{99}
\bibitem{HYR2} As for the $R^2$-gravity in three dimension, we will report
our results elsewhere.
\end{thebibliography}
\end{document}